\documentclass[prd,twocolumn,showpacs,preprintnumbers,amsmath,amssymb]{revtex4}

\usepackage{graphicx}

\usepackage{bm}

\newcommand{\beq}{\begin{equation}}
\newcommand{\eeq}{\end{equation}}
\newcommand{\beqs}{\begin{eqnarray}}
\newcommand{\eeqs}{\end{eqnarray}}

\begin{document}

\title{Constraints on $N_c$ in Extensions of the Standard Model} 

\author{Robert Shrock}

\affiliation{
C. N. Yang Institute for Theoretical Physics \\
State University of New York \\
Stony Brook, NY 11794}

\begin{abstract}

We consider a class of theories involving an extension of the Standard Model
gauge group to an {\it a priori} arbitrary number of colors, $N_c$, and derive
constraints on $N_c$.  One motivation for this is the string theory landscape.
For two natural classes of embeddings of this $N_c$-extended Standard Model in
a supersymmetric grand unified theory, we show that requiring unbroken
electromagnetic gauge invariance, asymptotic freedom of color, and three
generations of quarks and leptons forces one to choose $N_c=3$.  Similarly, we
show that for a theory combining the $N_c$-extended Standard Model with a
one-family SU(2)$_{TC}$ technicolor theory, only the value $N_c=3$ is allowed.

\end{abstract}


\maketitle

Although the Standard Model (SM), as augmented to allow for massive neutrinos
and lepton mixing, is in agreement with current particle physics data, there
are many basic questions that it leaves unanswered.  One of these is why the SM
gauge group $G_{SM} = {\rm SU}(3)_c \times {\rm SU}(2)_L \times {\rm U}(1)_Y$
has a color group involving $N_c=3$ colors instead of a different value.  Some
insight into this question is obtained from a grand unified theory (GUT) in
which $G_{SM}$ is embedded in a simple group, $G_{GUT}$.  But one may argue
that this just shifts the issue to a higher energy scale, where it becomes
the question of why nature picks a particular $G_{GUT}$ and why it is broken
to $G_{SM}$.  Indeed, recent studies in string theory suggest an exponentially
large number of vacua (the ``landscape'') which yield different physics below
the Planck scale, including different gauge groups and values of fundamental
constants and parameters \cite{landscape}-\cite{rank}.

Accordingly, we investigate here the following question: for a class of
theories (in $3+1$ dimensions, at zero temperature) involving the
$N_c$-extended Standard Model (ESM) gauge group, 
\beq
G_{ESM} = {\rm SU}(N_c) \times {\rm SU}(2)_L \times {\rm U}(1)_Y \ , 
\label{gesm}
\eeq
do reasonable criteria lead one to choose the observed value $N_c=3$?  This
study is motivated partly by the goal of applying selection criteria to choose
from among the vast set of {\it a priori} possible vacua in the string theory
landscape and partly by the general goal of understanding better the properties
of SM-like theories. We carry out our analysis first using two natural classes
of embeddings of $G_{ESM}$ in a grand unified theory.  Specifically, we explore
how well one can restrict $N_c$ by imposing only three conditions: (C1) exact
electromagnetic U(1)$_{em}$ gauge invariance, (C2) asymptotic freedom of the
SU($N_c$) color gauge interaction with resultant confinement of color, and (C3)
the observed number of SM fermion generations (families), $N_g=3$.  It will
turn out that for $N_c \ge 4$, our GUT constructions will lead to additional
matter fermions with SM quantum numbers, but at a minimum, we require $N_g=3$
families of the usual quarks and leptons.  A change in $N_c$ implies changes in
many other properties of a theory, such as baryon/meson mass ratios, hence
nuclear binding energies, etc.  We provisionally accept such changes and only
impose the three conditions above. A plausible additional condition could be to
require that $N_c$ is odd, so that baryons are fermions and nuclei exhibit the
usual shell structure.  However, interestingly, we find that the above
conditions (i)-(iii) are, by themselves, sufficient to determine $N_c$
\cite{largen}.

For our GUT study we require gauge coupling unification and hence assume the
requisite $N_c$-extension (symbolized with E) of the minimal supersymmetric
standard model (denoted MSESM) or split supersymmetry (denoted ESS); in the
latter, all scalars except the Higgs have masses comparable to the GUT-scale,
$M_{GUT}$, while fermionic superpartners have masses of order the electroweak
symmetry breaking (EWSB) scale \cite{splitsusy}.  In the $N_c=3$ special case,
these yield gauge coupling unification with the physical values of the SM gauge
couplings \cite{sgut}-\cite{gcunification}; for other values of $N_c$, we only
assume a GUT and allow the SM gauge couplings at $m_Z$ to vary from their
physical values.  As usual, we denote $\Lambda_{QCD}$ as the scale where the
SU($N_c$) gauge coupling $\alpha_c$ grows to O(1).  Since increasing $N_c$,
with other parameters held fixed, increases the rate of running of $\alpha_c$
and hence decreases the ratio $M_{GUT}/\Lambda_{QCD}$ and increases the proton
decay rate, we allow ourselves the freedom to decrease the common GUT gauge
coupling $\alpha_{GUT}$ from its conventional value of $\alpha_{GUT} \simeq
1/24$ to compensate for this effect.

We begin with GUT embeddings of $G_{ESM}$. Since the rank $rk(G_{ESM})=N_c+1$,
a minimal GUT can be constructed with the group $G_{GUT}={\rm SU}(N) \supset
G_{ESM}$, where
\beq
N=N_c+2 \ , 
\label{nnc}
\eeq
and hence $rk(G_{GUT})=rk(G_{ESM})$.  A natural generalization of the matter fermion
content of the SM and SU(5) GUT \cite{gg,sgut} which keeps the color
interaction asymptotically free for arbitrary $N_c$ is to assign the matter
fermions and corresponding chiral superfields to $N_g=3$ generations of the
anomaly-free set
\beq
[2]_N+(N-4)[\bar 1]_N \ , 
\label{gutfermions1}
\eeq
where $[k]_N$ denotes the antisymmetric rank-$k$ tensor representation of
SU($N$), with dimension ${\rm dim}([k]_N)={N \choose k}$, and we write 
fermion fields as left-handed.

The operators for weak hypercharge $Y$ and electric charge $Q=T_3+(Y/2)$ are
the $N \times N$ matrices \cite{lq,snsq} 
\beq
 Y={\rm diag}(-2/N_c..., -2/N_c,1,1)
\label{y}
\eeq
and
\beq
Q={\rm diag}(-1/N_c..., -1/N_c,1,0) \ .
\label{q}
\eeq
The SU($N$) GUT group is envisioned to be broken to $G_{ESM}$ at the GUT scale
$M_{GUT}$, while $G_{ESM}$ should be broken to
U(1)$_{em}$ at the electroweak scale. 

With respect to $G_{ESM}$, the matter fermions transform as $N-4$ copies of 
\beq
[\bar 1]_N: \ (\bar{N}_c,1)_{2/N_c} + (1,2)_{-1} \ , 
\label{1bardecomp}
\eeq
where the numbers denote the representations and subscripts
denote $Y$, with corresponding fields
\beq
d^c_{a,p,L} \ , \quad L_{p,L}={\nu_e \choose e}_{p,L} 
\label{1barfields}
\eeq
($p=1...N-4$ being a copy index), and
\beq
[2]_N: \quad ([2]_{N_c},1)_{-4/N_c} + (N_c,2)_{1-(2/N_c)} + (1,1)_2
\label{2decomp}
\eeq
with fields 
\beq
\xi^{ab}_L \ , \quad {u^a \choose d^a}_L \ , \quad e^c_L
\label{2fields}
\eeq
for the first generation, and similarly for the higher generations, where $1
\le a,b \le N_c$ are color indices.  

The electric charges of the color-nonsinglet matter fermions are $q_d = q_u-1 =
q_\xi/2 = -1/N_c$.  If and only if $N_c=3$, then $\xi_L$ is a $\bar 3$ and is
$u^c_L$; for larger $N_c$, it is a distinct (antisymmetric rank-2) 
representation, $[2]_{N_c}$. Since this GUT is a chiral gauge
theory, no fermion bare mass terms are present at the GUT scale.  For each
generation, this theory includes $N_d=2(N_c-1)$ SU(2)$_L$ doublets of matter
fermions, of which $N_c$ are color-nonsinglets and $N_c-2$ are color-singlets
(leptons).  We exclude the value $N_c=2$ because the resulting theory would not
have any leptonic SU(2)$_L$ doublets.

Below the GUT scale, with GUT-mass color-nonsinglet Higgs superfields
integrated out \cite{colorhiggs}, the leading coefficient of the SU($N_c$) beta
function \cite{beta} is
\beq
b_0^{(c)} = \begin{cases} 3N_c-N_g(N_c-1)     & \text{for MSESM} \cr
                    (1/3)[9N_c-2N_g(N_c-1)]    & \text{for ESS} \ . 
              \end{cases}
\label{b0cfirst}
\eeq
For both of these cases, with the physical value $N_g=3$,
the color SU($N_c$) gauge interaction is asymptotically free for all $N_c$, as
required by (C2).

Now the U(1)$_{em}$ gauge interaction is vectorial if and only if the charges
of the (left-handed) fermions can be written as a set of equal and opposite
pairs together with possible zero entries.  For our analysis it will suffice to
consider a single generation.  The charges of the fermions in the $[2]_N$ are
(i) $-2/N_c$, with multiplicity $\nu=N_c(N_c-1)/2$; (ii) $-(1/N_c)+1$ with
$\nu=N_c$; (iii) $-1/N_c$ with $\nu=N_c$; and (iv) 1 with $\nu=1$.  The charges
of the matter fermions in the $N-4=N_c-2$ copies of the $[\bar 1]_N$ are (v)
$1/N_c$ with $\nu=(N_c-2)N_c$; (vi) $-1$ with $\nu=N_c-2$; and (vii) 0 with
$\nu=N_c-2$.  Aside from the excluded case $N_c=2$, these charges consist of
equal and opposite pairs if and only if $N_c=3$.  Hence, for $N_c \ge 4$,
U(1)$_{em}$ is an (anomaly-free) chiral gauge interaction.  Since the color
gauge interaction is asymptotically free, $\alpha_c$ increases with decreasing
mass scale and eventually becomes large enough to produce bilinear fermion
condensates, some of which violate U(1)$_{em}$, thereby giving the photon a
mass.  This rules out such models and shows that, aside from the already
excluded $N_c=2$ case, the only physically allowed value of $N_c$ in this class
of models is $N_c=3$.

We illustrate this in the simplest case, $N_c=4$, i.e., $N=6$. Here the matter
fermions for a given generation transform as $[2]_6 + 2([\bar 1]_6)$, and
hence, with respect to $G_{ESM}$, as
\beq
(6,1)_{-1}+(4,2)_{1/2}+(1,1)_2+2(\bar 4,1)_{1/2}+2(1,2)_{-1}
\label{nc4reps}
\eeq
with fields given by (\ref{1barfields}) and (\ref{2fields}).  The
color-nonsinglets have charges $q_d=q_u-1=-1/4$ and $q_\xi=-1/2$. Note that
$[2]_4 \approx [\bar 2]_4$ (in general, $[k]_N \approx \overline{[N-k]}_N$).
As the mass scale decreases from large values and $\alpha_c$ increases, the
first bilinear matter fermion condensates to form are those in the most
attractive channel (MAC).  A measure of the attractiveness of a condensation
channel of the form $R_1 \times R_2 \to R_{cond.}$ is $\Delta C_2 =
C_2(R_1)+C_2(R_2)-C_2(R_{cond.})$, where $C_2(R)$ is the quadratic Casimir for
the representation $R$.  Since color is vectorial here, condensation channels
have $R_2 = \bar R_1$, $R_{cond.}=1$, and $\Delta C_2 = 2C_2(R_1)$.  The MAC is
\beq
(6,1)_{-1} \times (6,1)_{-1} \to (1,1)_{-2}
\label{661channel}
\eeq
with $\Delta C_2=2C_2([2]_4)=5$ and associated condensate $\langle
\epsilon_{abrs}\xi_L^{ab \ T} C \xi_L^{rs}\rangle$ with $q=y/2=-1$.  This
violates U(1)$_{em}$ (as well as U(1)$_Y$) and gives the photon a mass.
In passing, we note that for $N_c=5$, i.e., $N=7$, color itself becomes a
chiral gauge interaction and the MAC for color-nonsinglet matter fermions,
viz., $([2]_5,1)_{-4/5} \times ([2]_5,1)_{-4/5} \to ([\bar 1]_5,1)_{-8/5}$, 
breaks not only U(1)$_{em}$ and U(1)$_Y$, but also self-breaks 
color SU(5)$_c$ to SU(4)$_c$.  (The SU(4)$_c$ theory is vectorial and
does not break further.) 

For odd $N_c=2m-1$, a second choice for the left-handed matter fermion content
in the SU($N$) GUT with $N=N_c+2=2m+1$ is $N_g=3$ copies of the anomaly-free
set~\cite{g79}
\beq
\sum_{\ell=1}^m \ [2\ell]_N \ . 
\label{gutfermions2}
\eeq
Since 
$[N-k]_N \approx [\bar k]_N$, the $\ell=m$ term is $[N-1]_N \approx [\bar
1]_N$.  The set (\ref{gutfermions2}) is a natural generalization of the $N_c=3$
set $[2]_5+[4]_5 \approx [2]_5 + [\bar 1]_5$ for SU(5) \cite{gg,sgut}.  The
group SU($N$) with $N=2m+1$ has an embedding in SO($4m+2$) given by ${\rm
SU}(2m+1) \times {\rm U}(1)_X \subset {\rm SO}(4m+2)$, where U(1)$_X$ is an
additional U(1) symmetry.  The SO($M$) groups with $M=2$ mod 4 and $M \ge 10$
have complex representations but are anomaly-free.  The total number of chiral
matter fermions in (\ref{gutfermions2}) is
\beq
\sum_{\ell=1}^m {2m+1 \choose 2\ell} = 2^{2m}-1 \ . 
\label{nfermions}
\eeq
Adding an SU($N$)-singlet field to the set (\ref{gutfermions2}) thus 
yields $2^{2m}=2^{N_c+1}$ chiral fermions, which fit exactly in 
the spinor representation of SO($4m+2$).  
The color-nonsinglet
matter fermions in (\ref{gutfermions2}) comprise the vectorial set 
$2\sum_{k=1}^{m-1} \{ [k]_{N_c} + [\bar k]_{N_c} \}$.
We calculate that in the
interval between $\sim 1$ TeV and $M_{GUT}$, with the heavy color-nonsinglet
Higgs chiral superfields integrated out,
\beq
b^{(c)}_0 = \begin{cases} 3N_c-2^{N_c-2}N_g      & \text{for MSESM} \cr
                     (1/3)(9N_c-2^{N_c-1}N_g)    & \text{for ESS} \ .
             \end{cases}
\label{b0csecond}
\eeq
With $N_g=3$, it follows that for $N_c=3$, $b_0^{(c)}$ has the respective
values 3 and 5 for the MSSM and split SUSY, but we find that for $N_c \ge 5$, \
$b_0^{(c)} < 0$ for both the MSESM and ESS, i.e., color is non-asymptotically
free.  Thus, in this class of models, the constraint (C2) of asymptotic freedom
of color is satisfied only if $N_c=3$.  This conclusion is independent of
whether $Q$ is (proportional to) a generator of SU($N$), as in eq. (\ref{q}),
or $Q$ is a linear combination of generators of SU($N$) and U(1)$_X$ \cite{xy},
since this does not affect the embedding of color SU($N_c$) in SU($N$), so the
color SU($N_c$) representation content is the same for both choices.

If one were to require that the matter fermions of the $N_c$-extended SM
(together with an electroweak-singlet neutrino, $\nu^c$) of each generation fit
exactly in a spinor representation of SO($4m+2$), this would imply the
condition $4(N_c+1)=2^{N_c+1}$ \cite{nc}.  The only solution to this equation
is $N_c=3$.  A related result is that for the ESM or MSESM with just the quarks
and leptons of each generation, the cancellation of ${\rm SU}(2)_L^2 {\rm
U}(1)_Y$ and ${\rm U}(1)_Y^3$ anomalies for each generation occurs if and only
if $N_cY_Q+Y_L=0$, where $Y_Q$ and $Y_L$ denote the hypercharges of the quark
and lepton SU(2)$_L$ doublets.  This condition is equivalent to
$q_d=q_u-1=-(1/2)[1+N_c^{-1}(2q_e+1)]$.  With the usual choices
$q_e=q_\nu-1=-1$, this yields $q_d=q_u-1=(1/2)(-1+N_c^{-1})$ \cite{nc,tmyan}.
These values of $q_d$ and $q_u$ are equal to the values obtained from
eq. (\ref{q}) if and only if $N_c=3$.  In either of our GUT embeddings of the
ESM, the fermion content is such that there is no ${\rm SU}(2)_L^2 {\rm
U}(1)_Y$ or ${\rm U}(1)_Y^3$ gauge anomaly; for $N_c \ge 4$, this is due to
additional fermion contributions beyond those of the usual SM, such as the
$\xi$ fields and the $N_c-3$ additional copies of $[\bar 1]_N$.

So far we have worked in the context of a supersymmetric GUT with EWSB via
Higgs vacuum expectation values. We next investigate constraints on $N_c$ for
models containing the ESM, in which EWSB is due to the dynamical formation of 
bilinear fermion condensates of certain fermions that are nonsinglets under an
asymptotically free, vectorial gauge interaction called technicolor (TC) that
becomes strongly coupled at the TeV scale \cite{tc}.  Here the conditions
(C1)-(C3) above are satisfied, and we derive our constraint from the
requirement that the technicolor theory be asymptotically free, so that it
confines and its coupling grows large enough to produce the technifermion
condensate which is the source of EWSB. We take the technicolor gauge group to
be SU($N_{TC}$) and the technifermions to transform according to the
fundamental representation of SU($N_{TC}$) and to comprise one SM family ${U
\choose D}_L$, ${N \choose E}_L$, and $F^c_L$, $F=U,D,N,E$.  In order to give
SM fermions masses, TC is embedded in a larger theory, extended technicolor
(ETC) \cite{etc,tg}. A natural embedding of TC in an ETC theory uses a gauge
group SU($N_{ETC}$) and gauges the SM generational index, combining it with TC,
so that $N_{ETC}=N_g+N_{TC}$ \cite{od}.  Modern one-family (E)TC models (e.g.,
\cite{aps}) are motivated to use $N_{TC}=2$ because this value (a) minimizes
technifermion loop modifications of the $Z$ propagator, as measured by the $S$
parameter \cite{scalc}; (b) plausibly produces walking behavior (associated
with an approximate infrared fixed point \cite{wtc}), which is necessary in
order to obtain sufficiently large SM fermion masses, and (c) is required by
the mechanism to explain light neutrinos in TC \cite{ntlrs}.  Combining
$N_{TC}=2$ and $N_g=3$, one gets $N_{ETC}=5$.  Technicolor theories are subject
to severe constraints, in particular, from precision electroweak measurements.
Walking technicolor theories may be able to satisfy these constraints
\cite{scalc}, although this question is currently unsettled; for the purposes
of our present analysis, we tentatively assume that they do.  The number of
SU(2)$_L$ doublets in one-family TC is $N_d=N_{ETC}(N_c+1)$.  In order to avoid
a global $\pi_4$ anomaly in the SU(2)$_L$ theory \cite{pi4}, $N_d$ must be
even.  Given that one uses $N_{TC}=2$, so that $N_{ETC}$ is odd, this implies
that $N_c$ must also be odd (and $\ge 3$).  Including all SM-nonsinglet
fermions, one has $b^{(c)}_0 = (1/3)(11N_c - 4N_{ETC})$, so the asymptotic
freedom of color at the EW scale requires $N_c > 4N_{ETC}/11=20/11$.  This
condition is always satisfied.  One also requires that TC be asymptotically
free.  The leading coefficient in the TC beta function is
$b^{(TC)}_0=(1/3)(11N_{TC}-2N_{TF})$, where $N_{TF}=2(N_c+1)$ in the one-family
TC model.  Hence, the constraint $b^{(TC)}_0 > 0$ yields the upper bound $N_c <
(11/4)N_{TC}-1$, i.e., for $N_{TC}=2$, $N_c < 9/2$.  Given that $N_c$ must be
odd, the only solution to this inequality is $N_c=3$.

In summary, the recent suggestion from string theory of a huge landscape of
vacua, leading, among other things, to different low-energy ($E << M_{Pl}$)
gauge groups, motivates one to explore how physically reasonable properties
constrain the structure of these gauge groups.  Here we have investigated how
the conditions (C1)-(C3) above constrain theories with the $N_c$-extended
SM gauge group $G_{ESM}$ embedded in a GUT group. We find that for two natural
embeddings, these conditions force $N_c=3$.  With the chiral fermions
(\ref{gutfermions1}), condition (C1) alone suffices to yield this result, while
for the chiral fermion set (\ref{gutfermions2}), conditions (C2)+(C3) suffice.
For a one-family SU(2)$_{TC}$ technicolor model, the asymptotic freedom of
technicolor implies $N_c=3$.  Although these do not exhaust the possibilities
for physics beyond the SM, they can thus give an interesting insight into why
$N_c=3$ in our world.

This research was partially supported by the grant NSF-PHY-03-54776.


\begin{thebibliography}{99}
\bibitem{lq}
We keep the lepton charges fixed at their usual values. The ESM with variable
lepton charges is analyzed in \cite{nc}.
\bibitem{snsq}
Hence, $\sin^2\theta_W = {\rm Tr}(T_3^2)/{\rm Tr}(Q^2) = N_c/[2(N_c+1)]$ at
the GUT scale, $M_{GUT}$.  Although one could investigate restrictions due to
an associated change in the low-energy value of $\alpha_{em}$, we will not
need to do this to derive our constraints on $N_c$.
\bibitem{colorhiggs}
We assume that any color-nonsinglet Higgs chiral superfields have 
masses of order $M_{GUT}$; lighter masses could produce excessive proton decay.
\bibitem{beta}
Our notation for the beta function for a gauge group $G_j$ is 
$\beta_j = d \alpha_j/dt = - (2\pi)^{-1}\alpha_j^2 \Big ( 
b_0^{(j)} + b_1^{(j)} \alpha_j/(4\pi) + O(\alpha_j^2) \Big )$ where 
$\alpha_j =g_j^2/(4\pi)$ and $t=\ln \mu$.
\bibitem{od}
Here, $[G_{ESM},G_{ETC}]=0$.  The situation in a one-doublet TC model is 
more complicated because $[G_{ESM},G_{ETC}] \ne 0$, so ETC gauge bosons
carry color and/or charge; we do not consider such models here.
\end{thebibliography}
\end{document}